\documentclass[12pt]{article} 
\usepackage[latin1]{inputenc}
\usepackage[german,english]{babel}
\usepackage{amssymb}
\usepackage{subfigure}
\usepackage{epsfig}
\usepackage{epsf}
\usepackage{graphicx,psfrag}
\usepackage{longtable}
\usepackage{rotating}

\begin{document}

\textbf{\Large 2-Soliton-solution of the Novikov-Veselov equation}
\vspace{1 cm}

\textbf{J. Nickel\footnotemark[1], H. W.
Schürmann\footnotemark[1]}

\vspace{1 cm}

\footnotetext[1]{Department of Physics, University of Osnabr\"uck,
D-49069 Osnabr\"uck, Germany, corresponding author:
jnickel@uos.de}

\small{ Based on a superposition method recently proposed to obtain 1-solitary wave solutions of the KdV-Burgers
equation \cite{Yua2005}, we show that this method can also be used
to find a 2-solitary wave solution of the Novikov-Veselov equation. Thus, it seems that the method
of Yuanxi and Jiashi in general is not restricted to constructing 1-solitary wave solutions of nonlinear wave and evolution equations (NLWEEs).}\\

\textbf{KEY WORDS:} Linear superposition; solitary wave solution, Novikov-Veselov equation.\\
\textbf{PACS:} 02.30.Jr

\section{Introduction}\label{Intro}

We analyze the Novikov-Veselov equation (NV equation) given by Hu \cite{Hu1994} which is
another (2 + 1)-dimensional analog of the KdV equation besides the
well-known Kadomtsev-Petviashvili equation \cite{Cheng1990}. It
has relevance in nonlinear physics (in particular in inverse
scattering theory)
\cite{Tag,Cheng1990,Ath1991,Hu1994,Hu1996,Kon2004} and mathematics
(cf. e.g. \cite{Tai,Fer}). To obtain a 2-soliton solution of the
NV equation Hu has developed a nonlinear superposition formula
\cite[Eq. (5)]{Hu1994}. In the following we show that a
2-solitary wave solution of the NV equation can even be obtained
by a linear superposition method given by Yuanxi and Jiashi
\cite{Yua2005} that was proposed to find 1-solitary wave solutions
of nonlinear wave and evolution equations.

\section{2-soliton solution of the NV equation obtained by linear superposition}

We consider the following equations

\begin{eqnarray}
U_t + U_{xxx} + 3 \left(U^2\right)_x = 0,\label{NVMulti.f8}\\
U_t + U_{yyy} + 3 \left(U^2\right)_y = 0,\label{NVMulti.f9}\\
 2 U_t + U_{xxx} + U_{yyy} + 3 (U\: \partial_{y}^{-1} U_x)_{x} + 3
(U\: \partial_{x}^{-1} U_y)_{y} = 0.\label{NVMulti.f10}
\end{eqnarray}

Obviously, the KdV equation (\ref{NVMulti.f8}), (\ref{NVMulti.f9}) and the NV equation (\ref{NVMulti.f10})
are related: The linear terms of
Eq. (\ref{NVMulti.f10}) are equal to the superposition of those of
Eq. (\ref{NVMulti.f8}) and Eq. (\ref{NVMulti.f9}) and the
nonlinear terms of Eq. (\ref{NVMulti.f10}) are equal to the
superposition of those of Eqs. (\ref{NVMulti.f8}) and
(\ref{NVMulti.f9}) if traveling waves are considered with $\partial_x^{-1} = \partial_y^{-1}$.
Following the ideas of Yuanxi and Jiashi \cite{Yua2005} we construct the solutions
of Eq. (\ref{NVMulti.f10}) by linear superposition of those to Eq. (\ref{NVMulti.f8}) and Eq. (\ref{NVMulti.f9}). \\

According to a method described by Schürmann and Serov \cite{39}
we can evaluate the following 1-solitary wave solutions of Eqs.
(\ref{NVMulti.f8}), (\ref{NVMulti.f9})

\begin{eqnarray}
U(x,t) &=& \frac{c}{2\: k} \mathrm{sech}^2\left[\frac{1}{2} \sqrt{\frac{c}{k^3}} (k x - c t - y_0)\right],\label{NVMulti.f11}\\
U(y,t) &=& \frac{c}{2\: k} \mathrm{sech}^2\left[\frac{1}{2}
\sqrt{\frac{c}{k^3}} (k y - c t -
x_0)\right],\label{NVMulti.f12}
\end{eqnarray}

where $x_0$ and $y_0$ are arbitrary constants. Combining these
solutions and choosing $k = 1$, so that $\partial_x^{-1} =
\partial_y^{-1}$,

\begin{equation}\label{NVMulti.f13}
U(x,y,t) = \frac{c}{2} \mathrm{sech}^2\left[\frac{1}{2}
\sqrt{c}\:(x + y - c t)\right]
\end{equation}

is a 1- solitary wave solution of Eq. (\ref{NVMulti.f10}).\\
We tentatively write a 2-solitary wave solution according to

\begin{eqnarray}\label{NVMulti.f14}
U(x,y,t) &=& a\: \mathrm{sech}^2\left[\frac{1}{2} \sqrt{c}\:(k_1 x
+ p_1 y - c t)\right] +
b\: \mathrm{sech}^2\left[\frac{1}{2} \sqrt{c}\:(k_2 x + p_2 y - c t)\right],\nonumber\\
z_i &=& \frac{1}{2} \sqrt{c}\:(k_i x + p_i y - c t),\: i \in
\{1,2\},
\end{eqnarray}

with $c > 0$ and $a,\: b,\: k_1,\: k_2,\: p_1,\: p_2$ to be determined. Inserting Eq. (\ref{NVMulti.f14})
into Eq. (\ref{NVMulti.f10}) leads to

\begin{eqnarray}\label{NVMulti.f15}
&\frac{2\: a\:\sqrt{c}\: \mathrm{sech}^2 z_1 \mathrm{tanh} z_1}{4
k_1 k_2 p_1 p_2}& \left( - 2 c k_1 k_2 p_1 p_2 (k_1^3 + p_1^3 -2)
- 6 k_2 p_2 (2 a - c k_1 p_1) (k_1^3 + p_1^3) \mathrm{sech}^2 z_1\right)\nonumber\\
&- \frac{2\: a\:\sqrt{c}\: \mathrm{sech}^2 z_1 \mathrm{tanh} z_1}{4 k_1 k_2 p_1 p_2}& (6 b (k_2 p_1 + k_1 p_2) (k_1^2 k_2 + p_1^2 p_2)\: \textrm{sech}^2 z_2)\nonumber\\
&+ \frac{2\: b\:\sqrt{c}\: \mathrm{sech}^2 z_2 \mathrm{tanh}
z_2}{4 k_1 k_2 p_1 p_2}& \left( - 2 c k_1 k_2 p_1 p_2 (k_2^3 +
p_2^3 -2)
- 6 k_1 p_1 (2 b - c k_2 p_2) (k_2^3 + p_2^3) \mathrm{sech}^2 z_2\right)\nonumber\\
&- \frac{2\: b\:\sqrt{c}\: \mathrm{sech}^2 z_2 \mathrm{tanh}
z_2}{4 k_1 k_2 p_1 p_2}& (6 a (k_2 p_1 + k_1 p_2) (k_1 k_2^2 + p_1
p_2^2)\: \textrm{sech}^2 z_1) = 0.
\end{eqnarray}

Assuming $\frac{2\: a\:\sqrt{c}\: \mathrm{sech}^2 z_1 \mathrm{tanh} z_1}{4
k_1 k_2 p_1 p_2} \neq 0$ and $\frac{2\: b\:\sqrt{c}\: \mathrm{sech}^2 z_2 \mathrm{tanh}
z_2}{4 k_1 k_2 p_1 p_2} \neq 0$ and setting the
coefficients of $\mathrm{sech}^2 z_i$ equal to zero we obtain

\begin{eqnarray}\label{NVMulti.f18}
a &=& \lambda_1 \frac{c p_2 (p_2^3 -2)^{\frac{1}{3}}}{2 (p_2^3 -1)^{\frac{2}{3}}},
b = -\frac{1}{2} \lambda_2 c p_2 (p_2^3 - 2)^{\frac{1}{3}},
k_1 = \lambda_3 \left( 2 - \frac{p_2^3}{p_2^3 - 1}
\right)^{\frac{1}{3}},\\ k_2 &=& -\lambda_4 (p_2^3 - 2)^{\frac{1}{3}}, p_1 = \lambda_5 \frac{p_2}{(p_2^3 -
1)^{\frac{1}{3}}} \:\:\:\:\:\:\textrm{with}\nonumber
\end{eqnarray}


\begin{table}[h]
\begin{center}
\begin{tabular}{@{}c |c c c c c}
 &$\lambda_1$ & $\lambda_2$ & $\lambda_3$ & $\lambda_4$ & $\lambda_5$\\\hline
I &$1$ & $1$ & $1$ & $1$ & $1$\\
II & $1$ & $-(-1)^{\frac{1}{3}}$ & $(-1)^{\frac{2}{3}}$ & $-(-1)^{\frac{1}{3}}$ & $-(-1)^{\frac{1}{3}}$\\
III & $1$ & $(-1)^{\frac{2}{3}}$ & $-(-1)^{\frac{1}{3}}$ & $(-1)^{\frac{2}{3}}$ & $(-1)^{\frac{2}{3}}$\\
IV & $-(-1)^{\frac{1}{3}}$ & $1$ & $(-1)^{\frac{2}{3}}$ & $1$ & $(-1)^{\frac{2}{3}}$\\
V & $-(-1)^{\frac{1}{3}}$ & $-(-1)^{\frac{1}{3}}$ & $-(-1)^{\frac{1}{3}}$ & $-(-1)^{\frac{1}{3}}$ & $1$\\
VI & $-(-1)^{\frac{1}{3}}$ & $(-1)^{\frac{2}{3}}$ & $1$ & $(-1)^{\frac{2}{3}}$ & $-(-1)^{\frac{1}{3}}$\\
VII & $(-1)^{\frac{2}{3}}$ & $1$ & $-(-1)^{\frac{1}{3}}$ & $1$ & $-(-1)^{\frac{1}{3}}$\\
VIII & $(-1)^{\frac{2}{3}}$ & $-(-1)^{\frac{1}{3}}$ & $1$ & $-(-1)^{\frac{1}{3}}$ & $(-1)^{\frac{2}{3}}$\\
IX &$(-1)^{\frac{2}{3}}$ & $(-1)^{\frac{2}{3}}$ & $(-1)^{\frac{2}{3}}$ & $(-1)^{\frac{2}{3}}$ & 1
\end{tabular}
\end{center}
\end{table}

New solutions are given if $p_2 \in \mathbb{R}$ can be chosen so that $k_i$ and $p_1$ are real, $a$ and $b$ may be complex.
Subject to certain conditions some of these solutions are even physical (real and bounded) solutions. As an example the solution

\begin{eqnarray}\label{NVMulti.f17}
U(x,y,t) &=& \frac{c p_2 (p_2^3 - 2)^{1/3}}{2 (p_2^3 - 1)^{2/3}}
\mathrm{sech}^2\left(\frac{\sqrt{c} (p_2 y + (p_2^3 -2)^{1/3} x -
c (p_2^3 -1)^{1/3} t)}{2 (p_2^3 - 1)^{1/3}}\right) \nonumber\\
&-& \frac{c p_2 (p_2^3 -
2)^{1/3}}{2}\mathrm{sech}^2\left(\frac{\sqrt{c}}{2} (-p_2 y +
(p_2^3 - 2)^{1/3} x + c t)\right),\\
 & & p_2 > \sqrt[3]{2}
\end{eqnarray}

\begin{figure}[htbp]
\centering
\includegraphics[width=0.9\linewidth]{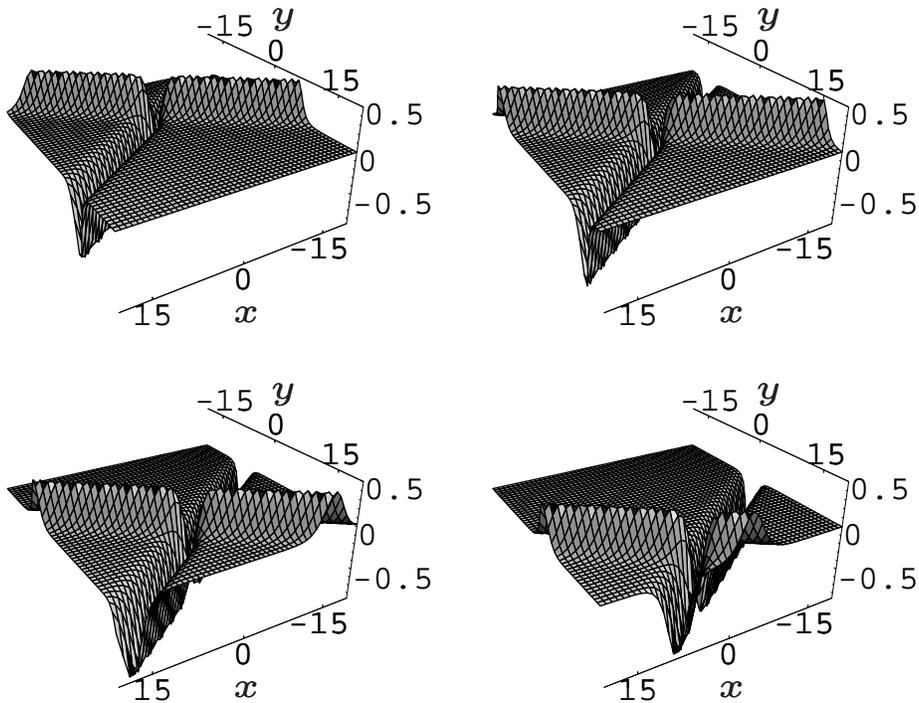}
\caption{\label{NVZweiSolitonYuanxi}\small 2-soliton solution (cf.
Eq. (\ref{NVMulti.f14}), I ) for $t = -10, 0,
10, 20$ ($c = 1$, $p_2 = 1.5$).}
\end{figure}

(according to Eqs. (\ref{NVMulti.f14}), (\ref{NVMulti.f18}) and line I of the above table) is shown in Fig. \ref{NVZweiSolitonYuanxi}. We have verified
this solution by putting it into the original equation
(\ref{NVMulti.f10}) utilizing Mathematica.

\section{Summary and concluding remarks}

By analyzing the structure of the NV equation we have shown that
by using a superposition method proposed to construct
1-solitary wave solutions of NLWEEs 2-solitary
wave solutions can be obtained. We suppose that the technique of
Yuanxi and Jiashi \cite{Yua2005} may lead to multi-solitary wave
solutions of certain NLWEEs if the NLWEE in quesion can be considered as a
superposition of NLWEEs that have the same type of solitary wave
solution (e.g., $\mathrm{sech}^2$); we leave this to future study.

\vspace{1 cm}

\section*{Acknowledgements}
The work was supported by the German Science Foundation (DFG)
(Graduate College 695 "Nonlinearities of optical materials").


\begin{thebibliography}{99}

\bibitem[Athorne {\em et al.} 1991]{Ath1991} Athorne C. and Nimmo J.~J.~C (1991), {\it Inverse Problems} \textbf{7}, 809-826.

\bibitem[Cheng 1990]{Cheng1990} Cheng Y. (1990), {\it J. Math. Phys.} \textbf{32}, 157-162.

\bibitem[Ferapontov 1999]{Fer} Ferapontov E.~V. (1999), {\it Differential
Geometry and its Applications} \textbf{11}, 117-128.

\bibitem[Hu 1994]{Hu1994} Hu X.-B. (1994), {\it J. Phys. A: Math. Gen.} \textbf{27}, 1331-1338.

\bibitem[Hu {\em et al.} 1996]{Hu1996} Hu X.-B. and Willox R. (1996), {\it J. Phys. A.: Math. Gen.} \textbf{29}, 4589-4592.

\bibitem[Konopelchenko {\em et al.} 1996]{Kon2004} Konopelchenko B. and Moro A. (2004), {\it J. Phys. A: Math. Gen.} \textbf{37}, L105-L111.

\bibitem[Schürmann {\em et al.} 2004]{39} Schürmann H. W. and Serov V. S. (2004), {\it Proc. Progress in Electromagnetics Research Symposium} March
28-31 (Pisa), 651-654.

\bibitem[Tagami 1989]{Tag} Tagami Y. (1989) arXiv: dg-ga/9511005 v5 20 Nov 1995.

\bibitem[Taimanov 1995]{Tai} Taimanov I.~A. (1995), {\it Phys. Lett. A} \textbf{141}, 116-120.

\bibitem[Yuanxi {\em et al.} 2005]{Yua2005} Yuanxi X. and Jiashi T. (2005),
{\it International Journal of Theoretical Physics} \textbf{44},
293-301.














\end{thebibliography}
\end{document}